\documentclass[twoside]{LCWS11}
\usepackage[latin1]{inputenc}
\usepackage[dvips]{graphicx,epsfig,color}
\usepackage{wrapfig,rotating}
\usepackage{amssymb,amsmath,array}
\usepackage{cite}
\def\tl{\tilde{t}_L}
\def\tr{\tilde{t}_R}

\def\beq{\begin{equation}}
\def\eeq#1{\label{#1}\end{equation}}
\def\eeqn{\end{equation}}
\def\beqa{\begin{eqnarray}}
\def\eeqa#1{\label{#1}\end{eqnarray}}
\def\eeqan{\end{eqnarray}}

\def\leqn#1{(\ref{#1})}

\begin{document}
\title{
SUSY-Yukawa Sum Rule at the LHC and the ILC} 
\author{Maxim Perelstein and Michael Saelim 
\thanks{This work is supported by the U.S. National Science Foundation through grant PHY-0757868 and CAREER award PHY-0844667.}
\vspace{.3cm}\\
Laboratory for Elementary Particle Physics, Cornell University \\
Ithaca, NY 14853, USA
}

\maketitle

\begin{abstract}
In this talk, we discuss the SUSY-Yukawa sum rule, a relation among masses and mixing angles of the third-generation squarks which follows directly from the coupling relation responsible for canceling the quadratic divergence in the Higgs mass. Radiative corrections modify the sum rule, introducing dependence on a variety of SUSY parameters beyond the third-generation squark sector. If some of those parameters are measured experimentally, a sharp prediction for the sum rule is possible. We demonstrate this point with a quantitative study. We also discuss the prospects for measuring the ingredients of the sum rule at the LHC, and argue that a high-energy $e^+e^-$ collider such as the ILC would be necessary to test the sum rule.  
\end{abstract}

\section{Introduction}

Supersymmetry (SUSY) is the only known mechanism which removes the quadratic divergence in the Higgs mass to all orders in perturbation theory, allowing the theory to remain perturbative to very high energy scales, such as GUT or Planck scale, without fine-tuning. SUSY predicts a number of new particles, with masses generically at the TeV scale, and searches for such particles are a major part of the LHC physics program. If the superparticles are within reach of the ILC, their masses and some of the couplings can be measured with high precision~\cite{ILC_RB}. The couplings between superpartners and the Higgs would be particularly important to measure, since they are unambiguously fixed by the 
requirement of quadratic divergence cancellation, and measuring them gives a unique test of the SUSY solution to the hierarchy problem. By the same token, these couplings are extremely model-independent: as long as the underlying mechanism of hierarchy stabilization is SUSY, they cannot be changed, while most other observables (superparticle spectrum, decay channels, etc.) depend on details of the SUSY model and breaking mechanism. The strongest of such couplings is the Higgs coupling to the superpartners of the top, the stop bosons, since it is related to the top Yukawa, the strongest coupling of the Standard Model Higgs. Can this coupling be measured, at the LHC or the ILC? On the one hand, there are reasons to be optimistic: naturalness suggests that stops must be rather light, ideally in the 300--400 GeV range, if SUSY is indeed responsible for stabilizing the Higgs~\cite{NSUSY1}. (Note also that, at the time of this writing, this mass range is not excluded by the LHC searches, as long as stops are significantly lighter than all other squarks~\cite{NSUSY1,NSUSY2}.) On the other hand, a direct measurement of the $hh{\tilde{t}}{\tilde{t}}$ vertex appears impossible, since the processes containing this vertex and quarks, gluons or electrons in the initial state have very small cross sections. The solution to this was proposed in Ref.~\cite{sumrule}, where a simple sum rule was formulated. The sum rule is a direct consequence of the SUSY relation between the top Yukawa and the $hh{\tilde{t}}{\tilde{t}}$ vertex, and at the same time it is made up entirely of potentially observable quantities, such as masses and mixing angles of the third-generation squarks. In this contribution, we will briefly describe the sum rule, and outline the prospects for testing it at the LHC~\cite{sumrule}. We will also describe how the LHC and ILC measurements of SUSY parameters outside of the third-generation squark sector can sharpen the theoretical prediction of the sum rule, by providing crucial information about the size of the radiative corrections to the sum rule. Finally, we will argue that an $e^+e^-$ collider, such as the ILC (or a higher-energy machine, if necessary to pair-produce stops and sbottoms) would be required to get sufficient information about the third-generation squark sector to unambiguously test the sum rule.

\section{SUSY-Yukawa Sum Rule}

The couplings of the Higgs to top and its partners, stops $\tl$ and $\tr$, have the form 
\beq
{\cal L} \,=\, \frac{y_t}{\sqrt{2}} h \bar{t} t + \frac{y_t^2}{2} h^2 \left( |\tl|^2+|\tr|^2\right)\,,
\eeq{lagr}
where $y_t$ is the top Yukawa constant. It is crucial for divergence cancellation, and guaranteed by SUSY, that the same $y_t$ appears in the two terms; the task is to test this fact experimentally. To our knowledge, a direct experimental measurement of the strength of the quartic interaction $hh{\tilde{t}}{\tilde{t}}$ is impossible. Once the Higgs gets a vev, $\left< h \right> = v$, a cubic interaction $h\tilde{t}\tilde{t}$ is generated, but it also seems very difficult to measure (although in some special cases this may be possible~\cite{Chang}). A mass term, $y_t^2 v^2 \left( |\tl|^2+|\tr|^2\right)$, is also generated, giving a mass precisely equal to $m_t$ to both stops. If this was the only contribution to the stop masses, it could be easily measured, providing a somewhat indirect but still very robust confirmation of the structure of Eq.~\leqn{lagr}. Of course, there are other contributions: the soft masses, $M_L^2 |\tl|^2+M_R^2|\tr|^2$, as well as the off-diagonal mass terms, $v(A_t \sin\beta -\mu \cos\beta) (\tl^*\tr+$ c.c.), and the D-term contribution. Nevertheless, it was shown in Ref.~\cite{sumrule} that the interesting contribution to the stop mass can be isolated and expressed in terms of physical observables. The SUSY prediction takes the form
\beq
m_t^2-m_b^2 = m_{t1}^2 \cos\theta_t + m_{t2}^2 \sin\theta_t -m_{b1}^2 \cos\theta_b-m_{b2}^2 \sin\theta_b-m_W^2 \cos2\beta\,,
\eeq{sumrule1} 
where $m_{ta}$, $m_{ba}$ are the physical stop and sbottom masses, respectively ($a=1, 2$), and $\theta_t$, $\theta_b$ are the rotation angles between the gauge and mass bases in the stop and sbottom sectors. This prediction was called "SUSY-Yukawa sum rule" in~\cite{sumrule}. It is convenient to define a dimensionless quantity
\beq
\Upsilon \equiv \frac{m_{t1}^2 \cos\theta_t + m_{t2}^2 \sin\theta_t -m_{b1}^2 \cos\theta_b-m_{b2}^2 \sin\theta_b}{v^2}\,.
\eeq{ups_def}
SUSY predicts (at the tree level, in the large $\tan\beta$ limit) 
\beq
\Upsilon^{\rm tree}_{\rm SUSY} = 0.28\,.
\eeq{Ups_tree}
By measuring stop and sbottom sector masses and mixing angles, a task that's difficult but may not be impossible as we discuss below, this prediction can be tested.

Before proceeding, let us discuss loop corrections to the prediction~\leqn{Ups_tree}. The masses in the definition of $\Upsilon$ are physical (pole) masses; one can also define the ``running" version of this observable, $\Upsilon(\mu)$, which has the same form but with masses and mixings taking their running values evaluated at scale $\mu$. The operations leading to the sum rule rely only on SUSY and $SU(2)_L$ gauge symmetry, so the tree-level sum rule applies to $\Upsilon(\mu)$ as long as $\mu$ is above SUSY and electroweak symmetry breaking scales. Thus, the only corrections to the sum rule are threshold effects, with no large logs. Numerically, however, these corrections can be large, since the sum rule involves a delicate cancellation among the stop and sbottom terms, and even fractionally small corrections to each term can result in significant fractional corrections in $\Upsilon$. This fact was already noted in Ref.~\cite{sumrule}, and 
will be further illustrated by the numerical work in the next section. This appears to diminish the usefulness of the sum rule. 
However, the large radiative correction is troublesome only if it is unknown; if it can be calculated and subtracted, the sum rule can still be meaningfully tested. Calculating the radiative corrections to the stop and sbottom masses requires knowledge of SUSY parameters, such as, for example, the gluino and chargino masses. In the next section, we show that experimental measurements of these masses at the LHC and ILC can significantly reduce the uncertainty on the theoretical prediction of $\Upsilon$.

\section{Improving the Theoretical Prediction of the Sum Rule with Data}

While at tree level SUSY prediction for $\Upsilon$ is just a fixed number (with only a slight $\tan\beta$ dependence),
radiative corrections to $\Upsilon$ depend on a number of SUSY parameters. If these parameters are treated as unknown, SUSY prediction for $\Upsilon$ is significantly washed out, and a broad range of values is possible (see Fig.~2 in Ref.~\cite{sumrule}). However, experimental measurement of SUSY parameters should clearly shrink this range. Testing the sum rule can then be thought of as consisting of two steps: (a) measure as many parameters as possible {\it not including third-generation squark masses and mixings}, and use them to narrow the range of radiative corrections to $\Upsilon$; and (b) measure the third-generation squark masses and mixings, and check that the combination in Eq.~\leqn{ups_def} falls within the range determined in (a). In this section, we present a Monte Carlo study of the step (a) of this procedure. 

Our study is in the context of the phenomenological MSSM (pMSSM)~\cite{pMSSM}. We assumed that the ``correct" model is the well-known benchmark point LCC1. We then scanned the pMSSM parameter space, and recorded the values of $\Upsilon$ at each point. To make efficient use of computing time, we only scan the parameters that significantly affect 
$\Upsilon$, namely: $M_1$, $M_2$, $M_3$, $m_{\tilde{Q}}$, $m_{\tilde{t}_R}$, $m_{\tilde{b}_R}$, $A_t$, $A_b$, $M_A$ (pole), $\tan\beta$ ($m_Z$), and $\mu$. We fix all other parameters at their LCC1 values. Even with this simplification, 
the traditional method of scanning over a grid is computationally prohibitive, leading us to use Markov Chain Monte Carlo (MCMC) techniques as detailed in~\cite{Peskin}. The MCMC algorithm is implemented in C++ with the GNU Scientific Library, and interfaces with SuSpect~\cite{SuSpect} for all the MSSM spectrum calculations.  As in~\cite{Peskin}, we initialize 50 Markov chains around the benchmark point LCC1, propagate them for one million steps, burn the first 10\% of each chain, and test for convergence of the algorithm with the Fourier analysis detailed in~\cite{Dunkley}. 

\begin{wrapfigure}{r}{0.5\columnwidth}
\centerline{\includegraphics[width=0.45\columnwidth]{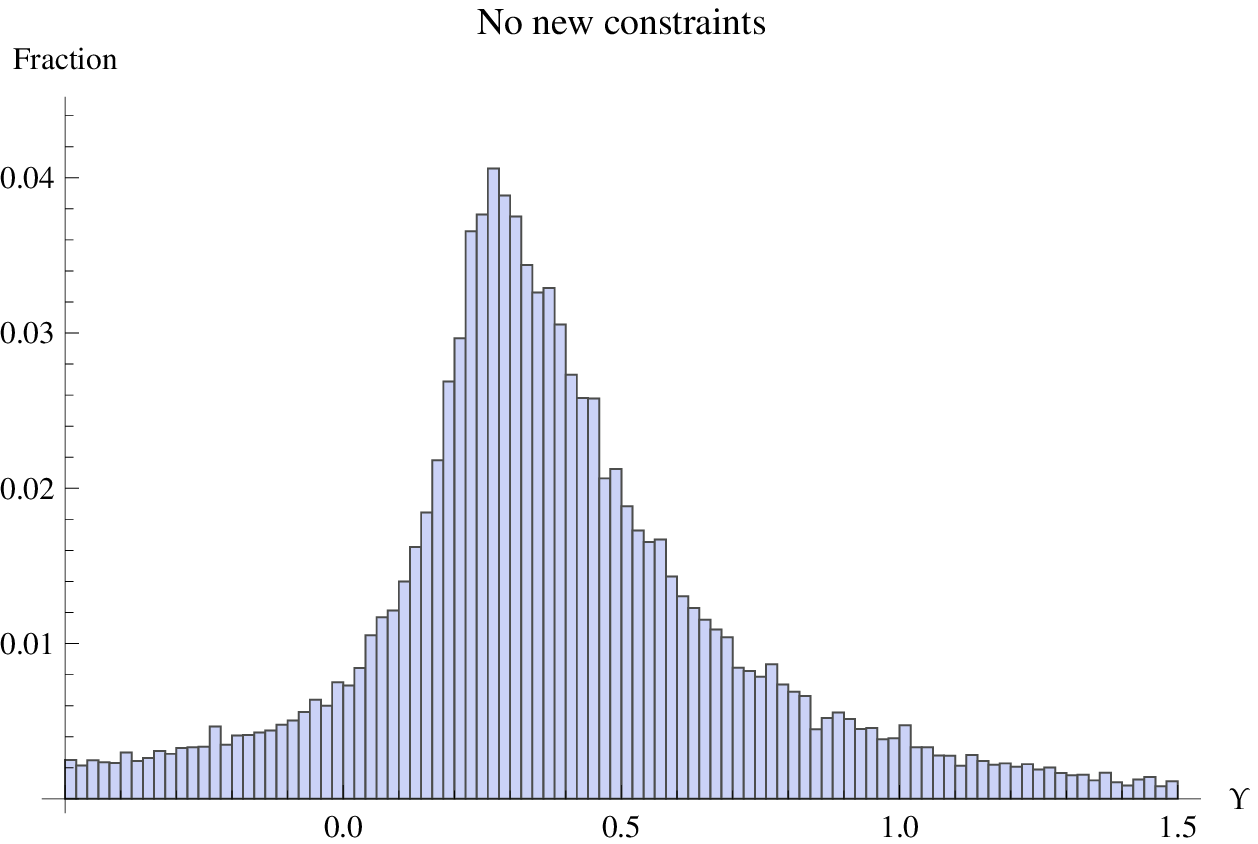}}
\vspace{0.4cm}
\centerline{\includegraphics[width=0.45\columnwidth]{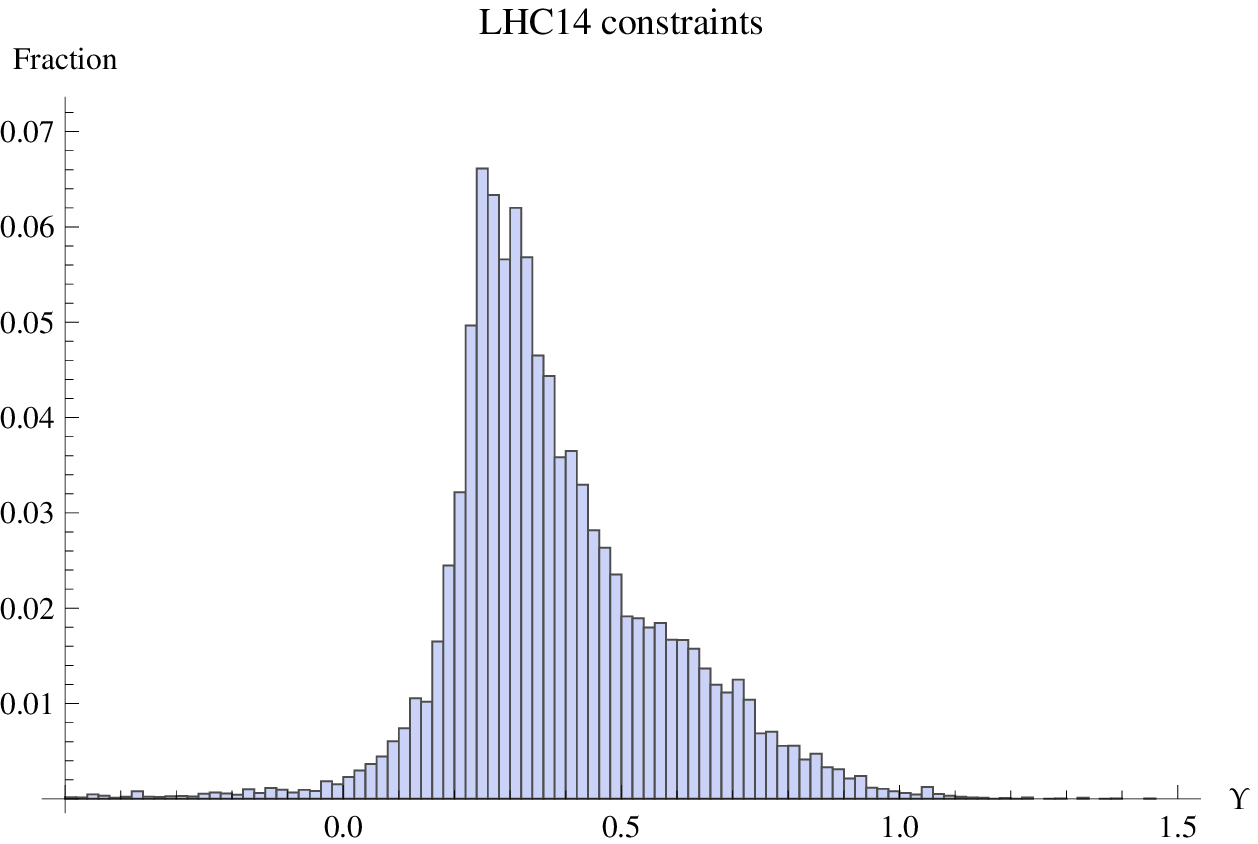}}
\vspace{0.4cm}
\centerline{\includegraphics[width=0.45\columnwidth]{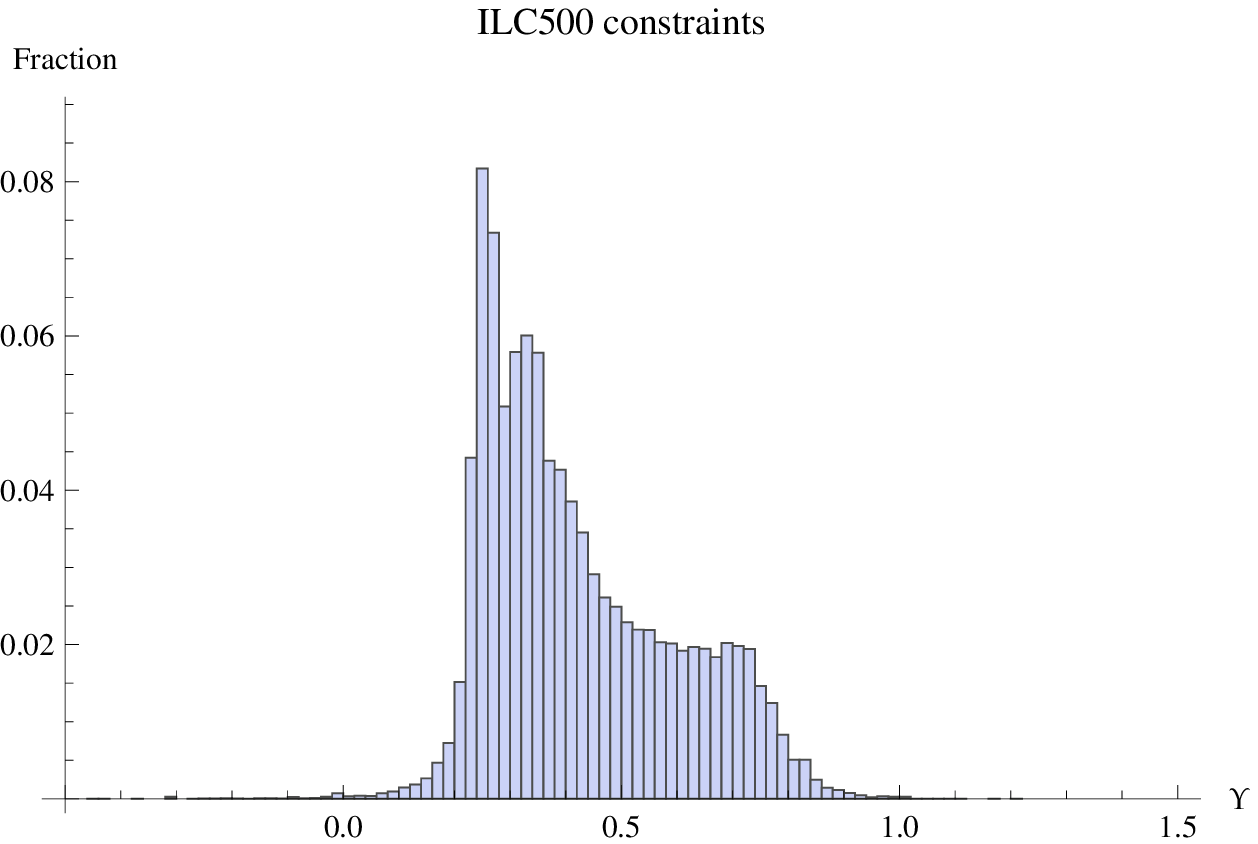}}
\caption{Distributions of $\Upsilon$ with different sets of experimental constraints. The assumed true model is the MSSM at the benchmark point LCC1.}\label{Fig:UpsilonDistribution}
\end{wrapfigure}

The first scan does not assume any experimental knowledge of the superpartner masses, beyond the requirement of a neutralino LSP, the LEP constraints on charged superpartner masses ($>100$ GeV) and the lightest CP-even Higgs mass (we use $m_h>108$ GeV, to conservatively take into account the uncertainty of the SuSpect prediction), as well as the current experimental constraints on $m_W$, $g_\mu - 2$, and Br($b\rightarrow s\gamma$). (We do not take into account the dark matter relic density constraint, which is subject to model-dependent cosmological uncertainties.) The top panel of Fig.~\ref{Fig:UpsilonDistribution} shows the resulting distributions of $\Upsilon$. We then repeated the scan with additional constraints on the SUSY parameters from measurements at the LHC-14 (middle panel) and the ILC-500 (bottom panel). The estimates of the uncertainties in the LHC and ILC measurements are taken from the 2004 report of the LHC/LC study group~\cite{LHCLC} (for a concise summary of these estimates, see Table 2 of Ref.~\cite{Peskin}). In all cases, we ignore information about third generation squarks, as explained above.

Figure~\ref{Fig:UpsilonDistribution} clearly demonstrates that, as expected, the ``theoretical" prediction of $\Upsilon$ becomes sharper as more information on SUSY spectrum is gathered, allowing to nail down the radiative corrections to stop and sbottom masses. It is useful to quantify this by computing the mean and standard deviation of the $\Upsilon$ predictions. Without any new experimental constraints from the LHC or ILC, $\Upsilon = 0.18 \pm 0.85$.  After the LHC-14 measurements, it narrows to $\Upsilon = 0.37 \pm 0.39$, while the ILC measurements at $\sqrt{s} = 500\,\textrm{GeV}$ narrow it further to $\Upsilon = 0.42 \pm 0.19$. (The true value at LCC1 is $\Upsilon = 0.27$.)

Note that, at the time of this writing, the point LCC1 is already ruled out by the LHC data. It is clear that qualitative lessons of this study apply throughout the model parameter space, although of course the amount of information a given collider can obtain does depend strongly on the spectrum. 

\section{Mass and Mixing Angle Measurements}

To test the sum rule, the theoretical prediction discussed above must be compared to $\Upsilon$ computed by directly 
measuring stop and sbottom masses and mixing angles. Ref.~\cite{sumrule} studied the potential for mass measurements at the LHC. The analysis relied on direct stop decays, $\tilde{t}\to t \tilde{\chi}^0_1$, and cascade decays of gluinos via sbottoms, $\tilde{g}\to \tilde{b}b$, $\tilde{b}\to b \tilde{\chi}^0_1$. Using the kinematic edge, as well as recently developed techniques such as subsystem $M_{T2}$ variables~\cite{MT2sub}, it was possible to extract the lighter stop and sbottom masses to roughly 10\% accuracy at the benchmark point used in this study. Recently, a more refined version of the gluino cascade analysis has been performed by D.~Curtin~\cite{DC}, confirming this conclusion for the sbottom mass. (Note that the benchmark point used in these studies has precisely the sort of spectrum favored by current LHC constraints and naturalness, with only third-generation squarks, gluino and the LSP appearing below 1 TeV~\cite{NSUSY1}.) Thus, it appears that the LHC can do a decent job on measuring the masses, at least as long as substantial samples of stops and sbottoms are produced, either directly or in cascade decays. However, as already emphasized in the discussion of radiative corrections above, the sum rule involves a delicate cancellation between the stop and sbottom terms, and even fractionally small corrections to each term can result in significant fractional corrections in $\Upsilon$. It is absolutely crucial to measure the stop and sbottom masses {\it as precisely as possible}. An $e^+e^-$ collider, with sufficiently high center-of-mass energy to pair-produce the stops and sbottoms, would be an ideal instrument for this task.

While a lot of work has been done on superpartner mass measurements, measuring mixing angles has not attracted the same attention. There are several proposals in the literature for measuring the stop mixing angle~\cite{Nojiri,PW,JS,RTMP}. For example, Ref.~\cite{PW} proposed using the polarization of the top quarks produced in the direct decay $\tilde{t}\to t\tilde{\chi}^0_1$ as a handle on the mixing angle at the LHC; however, the measurement is quite challenging 
experimentally, and even if it could be done, additional information on the neutralino composition (bino, wino and higgsino fractions) would be required for this approach to succeed. There are, to our knowledge, no proposals for measuring the sbottom mixing angle at the LHC. Unless a way to do it is found, no test of the sum rule is possible at the LHC. An $e^+e^-$ collider, on the other hand, is ideally suited for measuring mixing angles. Stops and sbottoms are produced in $e^+e^-$ collisions via photon or $Z$ exchange. Since the $Z$ couples with different strengths to left- and right-handed squarks, the $Z$ couplings to physical squark states (mass eigenstates) depend explicitly on the mixing angles. For example, the coupling $Z \tilde{t}_1^* \tilde{t}_1$ has the form  
\beq
{\cal L} = -i e \left[ \left( \frac{1}{6} t_w - \frac{1}{2} t_w^{-1}\right) \cos^2\theta_t \,+\, \frac{2}{3}t_w \sin^2\theta_t \right] \,\tilde{t}^*_1 \partial_\mu \tilde{t}_1 Z^\mu\,,
\eeq{Zcoup}
where $t_w\equiv \tan\theta_w$. A measurement of the total stop or sbottom pair-production cross section gives a direct measurement of the mixing angles. This technique was explored in Ref.~\cite{stopmix}, where it was found that a rather precise determination of the stop mixing angle (fractional error of about 10\% on $\cos\theta_t$) was possible at a 500 GeV $e^+e^-$ collider. Beam polarization was found to play a crucial role in this measurement. These conclusions seem rather robust, and should be valid as long as the center-of-mass energy is high enough to produce $\tilde{t}_1$ pairs. 

\section{Conclusions}

In this contribution, we described the SUSY-Yukawa sum rule, a simple prediction of SUSY which follows directly from the crucial coupling relation responsible for canceling the quadratic divergence in the Higgs mass. The sum rule involves only directly observable quantities, {\it i.e.} masses and mixing angles of third-generation squarks. Radiative corrections to the sum rule depend on a number of other SUSY parameters. We showed how measuring those parameters at the LHC and the ILC can lead to a sharper theoretical prediction of the sum rule. We also discussed the prospects of measuring the third-generation squark masses and mixing angles experimentally. While the LHC can measure masses, the precision seems insufficient for a meaningful test of the sum rule. Moreover, mixing angle measurements at the LHC appear very difficult or impossible. An $e^+e^-$ collider such as the ILC can provide precise mass and mixing angle measurements, as long as the center-of-mass energy is sufficiently high to produce both stop and sbottom states. If SUSY-like new physics is found at the LHC, this set of measurements could form an important part of the physics case for the ILC or a higher-energy $e^+e^-$ collider.

\section{Acknowledgements}

We are grateful to David Curtin for useful discussions. MP is grateful to David Curtin and Monika Blanke for collaborating on Ref.~\cite{sumrule}.


\begin{footnotesize}


\end{footnotesize}


\end{document}